\documentclass[italian,english]{article}
\usepackage[latin9]{inputenc}
\usepackage{color}
\usepackage{graphicx}
\usepackage{amssymb}

\makeatletter
\newcommand{\lyxaddress}[1]{
\par {\raggedright #1
\vspace{1.4em}
\noindent\par}
}

\makeatother

\usepackage{babel}

\begin{document}

\title{\textbf{A longitudinal component in massive gravitational waves arising
from a bimetric theory of gravity}}

\author{\textbf{Christian Corda}}

\maketitle

\lyxaddress{\begin{center}
INFN - Sezione di Pisa and Università di Pisa, Via F. Buonarroti
2, I - 56127 PISA, Italy
\par\end{center}}

\lyxaddress{\begin{center}
\textit{E-mail address:} \textcolor{blue}{christian.corda@ego-gw.it} 
\par\end{center}}

\begin{abstract}
After a brief review of the work of de Paula, Miranda and Marinho
on massive gravitational waves arising from a bimetric theory of gravity,
in this paper it is shown that the presence of the mass generates
a longitudinal component in a particular polarization of the wave.
The effect of this polarization on test masses is performed using
the geodesic deviation. A the end of this paper the detectability
of this particular polarization is also discussed, showing that its
angular dependence could, in principle, discriminate such polarization
with respect the two ones of general relativity, if present or future
detectors will achieve a high sensitivity.
\end{abstract}

\section{Introduction}

The design and construction of a number of sensitive detectors for
gravitational waves (GWs) is underway today. There are some laser
interferometers like the VIRGO detector, being built in Cascina, near
Pisa by a joint Italian-French collaboration \cite{key-1,key-2},
the GEO 600 detector, being built in Hannover, Germany by a joint
Anglo-Germany collaboration \cite{key-3,key-4}, the two LIGO detectors,
being built in the United States (one in Hanford, Washington and the
other in Livingston, Louisiana) by a joint Caltech-Mit collaboration
\cite{key-5,key-6}, and the TAMA 300 detector, being built near Tokyo,
Japan \cite{key-7,key-8}. There are many bar detectors currently
in operation too, and several interferometers and bars are in a phase
of planning and proposal stages.

The results of these detectors will have a fundamental impact on astrophysics
and gravitation physics. There will be lots of experimental data to
be analyzed, and theorists will be forced to interact with lots of
experiments and data analysts to extract the physics from the data
stream.

Detectors for GWs will also be important to confirm or ruling out
the physical consistency of general relativity or of any other theory
of gravitation \cite{key-9,key-10,key-11,key-12}. This is because,
in the context of extended theories of gravity, some differences from
general relativity and the others theories can be seen starting by
the linearized theory of gravity \cite{key-9,key-10,key-12}. 

Some papers in the literature have also shown that a massive component
of gravitational waves could in principle be present in alternative
theories of gravity, like scalar-tensor theories \cite{key-12,key-13,key-14},
high-order theories \cite{key-15} and a bimetric theory \cite{key-16}.

Focusing our attention on this bimetric theory, after a brief review
of the work of de Paula, Miranda and Marinho on massive gravitational
waves from such as theory \cite{key-16}, which is due to provide
a context to bring out the relevance of the results, in this paper
it is shown that the presence of the mass generates a longitudinal
component in a particular polarization of the wave. The effect of
this polarization on test masses is performed using the geodesic deviation.
A the end of this paper the detectability of this particular polarization
is also discussed, showing that its angular dependence could, in principle,
discriminate such polarization with respect the two ones of general
relativity, if present or future detectors will achieve a high sensitivity.

\section{A review of massive gravitational waves from the bimetric theory}

An extension of linearized general relativity which takes into account
massive gravitons gives a weak-field stress-energy tensor \cite{key-16}

\begin{equation}
T_{\mu\nu}^{(m)}=-\frac{m_{g}}{8\pi}\{h_{\mu\nu}-\frac{1}{2}[(g_{0}^{-1})^{\alpha\beta}h_{\alpha\beta}](g_{0})_{\mu\nu}\},\label{eq: TEI}\end{equation}

where $m_{g}$ is the mass of the graviton, and $(g_{0})_{\mu\nu}$
the non-dynamical background metric (note: differently from \cite{key-16}
in this paper we work with $G=1$, $c=1$ and $\hbar=1$). In this
way the field equations can be obtained in an einstenian form like

\begin{equation}
G_{\mu\nu}=-8\pi(T_{\mu\nu}+T_{\mu\nu}^{(m)}),\label{eq: einstein}\end{equation}

where $T_{\mu\nu}$ is the ordinary stress-energy tensor of the matter.
General relativity is recovered in the limit $m_{g}\rightarrow0$.

Calling $g{}_{\mu\nu}$ the dynamic metric and putting 

\begin{equation}
g_{\mu\nu}=\eta_{\mu\nu}+h_{\mu\nu}\label{eq: linearizza}\end{equation}

with $|h_{\mu\nu}|\ll1$ equation (\ref{eq: einstein}) can be linearized
in vacuum (i.e. $T_{\mu\nu}=0$) obtaining

\begin{equation}
{}[]\overline{h}_{\mu\nu}=m^{2}\overline{h}_{\mu\nu},\label{eq: linearizzata}\end{equation}

where $[]$ is the d'Alembertian operator and $\overline{h}_{\mu\nu}\equiv h_{\mu\nu}-\frac{h}{2}\eta_{\mu\nu}.$

The general solution of this equation is \cite{key-16}

\begin{equation}
\overline{h}_{\mu\nu}=e_{\mu\nu}\exp(ik^{\alpha}x_{\alpha}),\label{eq: sol S}\end{equation}

where $e_{\mu\nu}$ is the polarization tensor. 

The condition of normalization $k^{\alpha}k_{\alpha}=m^{2}$ gives
$k=\sqrt{\omega^{2}-m^{2}}$ and a speed of propagation 

\begin{equation}
v(\omega)=\frac{\sqrt{\omega^{2}-m^{2}}}{\omega},\label{eq: velocita' di gruppo 2}\end{equation}

which is exactly the velocity of a massive particle with mass $m$
(it is also the group-velocity of a wave-packet \cite{key-12,key-14,key-15}).

Thus, assuming that the wave is propagating in the $z$ direction,
the metric perturbation (\ref{eq: sol S}) can be rewritten like 

\begin{equation}
\overline{h}_{\mu\nu}=e_{\mu\nu}\exp(ikz-i\omega).\label{eq: sol S 2}\end{equation}

Using a tetrade formalism, the authors of \cite{key-16} found six
independent polarizations states (see equations 28-33 of \cite{key-16}).
In the following section it will be shown that, from the polarization
labelled with $\Phi_{22}$ in \cite{key-16} (equations 32 and 38
of \cite{key-16}), a longitudinal force is present.

\section{The origin of a longitudinal component}

Let us consider equation 38 of \cite{key-16}. Putting $h_{g}\equiv h_{00}+h_{33}$,
this equation can be rewritten as 

\begin{equation}
\Phi_{22}=\frac{1}{8}h_{g}(t,z).\label{eq: hg}\end{equation}

Taken in to account only the $\Phi_{22}$ polarization in equation
(\ref{eq: sol S}) one gets\begin{equation}
\overline{h}_{\mu\nu}(t,z)=\frac{1}{8}h_{g}(t,z)\eta_{\mu\nu}\label{eq: perturbazione scalare}\end{equation}
and the corrispondent line element is the conformally flat one

\begin{equation}
ds^{2}=[1+\frac{1}{8}h_{g}(t,z)](-dt^{2}+dz^{2}+dx^{2}+dy^{2}).\label{eq: metrica puramente scalare}\end{equation}
Because the analysis on the motion of test masses is performed in
a laboratory environment on Earth, the coordinate system in which
the space-time is locally flat is typically used and the distance
between any two points is given simply by the difference in their
coordinates in the sense of Newtonian physics \cite{key-12,key-13,key-14,key-15,key-16,key-17,key-18}.
This frame is the proper reference frame of a local observer, located
for example in the position of the beam splitter of an interferometer.
In this frame gravitational waves manifest themself by exerting tidal
forces on the masses (the mirror and the beam-splitter in the case
of an interferometer). A detailed analysis of the frame of the local
observer is given in ref. \cite{key-17}, sect. 13.6. Here only the
more important features of this coordinate system are recalled:

the time coordinate $x_{0}$ is the proper time of the observer O;

spatial axes are centered in O;

in the special case of zero acceleration and zero rotation the spatial
coordinates $x_{j}$ are the proper distances along the axes and the
frame of the local observer reduces to a local Lorentz frame: in this
case the line element reads \cite{key-17}

\begin{equation}
ds^{2}=-(dx^{0})^{2}+\delta_{ij}dx^{i}dx^{j}+O(|x^{j}|^{2})dx^{\alpha}dx^{\beta}.\label{eq: metrica local lorentz}\end{equation}

The effect of the gravitational wave on test masses is described by
the equation

\begin{equation}
\ddot{x^{i}}=-\widetilde{R}_{0k0}^{i}x^{k},\label{eq: deviazione geodetiche}\end{equation}
which is the equation for geodesic deviation in this frame.

Thus, to study the effect of the massive gravitational wave on test
masses, $\widetilde{R}_{0k0}^{i}$ has to be computed in the proper
reference frame of the local observer. But, because the linearized
Riemann tensor $\widetilde{R}_{\mu\nu\rho\sigma}$ is invariant under
gauge transformations \cite{key-12,key-15,key-17}, it can be directly
computed from eq. (\ref{eq: perturbazione scalare}). 

From \cite{key-17} it is:

\begin{equation}
\widetilde{R}_{\mu\nu\rho\sigma}=\frac{1}{2}\{\partial_{\mu}\partial_{\beta}h_{\alpha\nu}+\partial_{\nu}\partial_{\alpha}h_{\mu\beta}-\partial_{\alpha}\partial_{\beta}h_{\mu\nu}-\partial_{\mu}\partial_{\nu}h_{\alpha\beta}\},\label{eq: riemann lineare}\end{equation}

that, in the case eq. (\ref{eq: perturbazione scalare}), begins

\begin{equation}
\widetilde{R}_{0\gamma0}^{\alpha}=\frac{1}{16}\{\partial^{\alpha}\partial_{0}h_{g}\eta_{0\gamma}+\partial_{0}\partial_{\gamma}h_{g}\delta_{0}^{\alpha}-\partial^{\alpha}\partial_{\gamma}h_{g}\eta_{00}-\partial_{0}\partial_{0}h_{g}\delta_{\gamma}^{\alpha}\};\label{eq: riemann lin scalare}\end{equation}
the different elements are (only the non zero ones will be written):

\begin{equation}
\partial^{\alpha}\partial_{0}h_{g}\eta_{0\gamma}=\left\{ \begin{array}{ccc}
\partial_{t}^{2}h_{g} & for & \alpha=\gamma=0\\
\\-\partial_{z}\partial_{t}h_{g} & for & \alpha=3;\gamma=0\end{array}\right\} \label{eq: calcoli}\end{equation}

\begin{equation}
\partial_{0}\partial_{\gamma}h_{g}\delta_{0}^{\alpha}=\left\{ \begin{array}{ccc}
\partial_{t}^{2}h_{g} & for & \alpha=\gamma=0\\
\\\partial_{t}\partial_{z}h_{g} & for & \alpha=0;\gamma=3\end{array}\right\} \label{eq: calcoli2}\end{equation}

\begin{equation}
-\partial^{\alpha}\partial_{\gamma}h_{g}\eta_{00}=\partial^{\alpha}\partial_{\gamma}h_{g}=\left\{ \begin{array}{ccc}
-\partial_{t}^{2}h_{g} & for & \alpha=\gamma=0\\
\\\partial_{z}^{2}h_{g} & for & \alpha=\gamma=3\\
\\-\partial_{t}\partial_{z}h_{g} & for & \alpha=0;\gamma=3\\
\\\partial_{z}\partial_{t}h_{g} & for & \alpha=3;\gamma=0\end{array}\right\} \label{eq: calcoli3}\end{equation}

\begin{equation}
-\partial_{0}\partial_{0}h_{g}\delta_{\gamma}^{\alpha}=\begin{array}{ccc}
-\partial_{z}^{2}h_{g} & for & \alpha=\gamma\end{array}.\label{eq: calcoli4}\end{equation}

Now, putting these results in eq. (\ref{eq: riemann lin scalare})
one obtains:

\begin{equation}
\begin{array}{c}
\widetilde{R}_{010}^{1}=-\frac{1}{16}\ddot{h}_{g}\\
\\\widetilde{R}_{010}^{2}=-\frac{1}{16}\ddot{h}_{g}\\
\\\widetilde{R}_{030}^{3}=\frac{1}{16}[]h_{g}.\end{array}\label{eq: componenti riemann}\end{equation}

But, putting the field equation (\ref{eq: linearizzata}) in the third
of eqs. (\ref{eq: componenti riemann}) it is

\begin{equation}
\widetilde{R}_{030}^{3}=\frac{1}{16}m^{2}h_{g},\label{eq: terza riemann}\end{equation}

which shows that the field is not transversal. 

Infact, using eq. (\ref{eq: deviazione geodetiche}) it results

\begin{equation}
\ddot{x}=\frac{1}{16}\ddot{h}_{g}x,\label{eq: accelerazione mareale lungo x}\end{equation}

\begin{equation}
\ddot{y}=\frac{1}{16}\ddot{h}_{g}y\label{eq: accelerazione mareale lungo y}\end{equation}

and 

\begin{equation}
\ddot{z}=-\frac{1}{16}m^{2}h_{g}(t,z)z.\label{eq: accelerazione mareale lungo z}\end{equation}

Then the effect of the mass is the generation of a \textit{longitudinal}
force (in addition to the transverse one).

\section{Geodesic deviation}

For a better understanding of this longitudinal force, let us analyse
the effect on test masses in the context of the geodesic deviation.

Following \cite{key-14} one puts\begin{equation}
\widetilde{R}_{0j0}^{i}=\frac{1}{16}\left(\begin{array}{ccc}
-\partial_{t}^{2} & 0 & 0\\
0 & -\partial_{t}^{2} & 0\\
0 & 0 & m^{2}\end{array}\right)h_{g}(t,z)=-\frac{1}{16}T_{ij}\partial_{t}^{2}h_{g}+\frac{1}{16}L_{ij}m^{2}h_{g}.\label{eq: eqs}\end{equation}

Here the transverse projector with respect to the direction of propagation
of the GW $\widehat{n}$, defined by

\begin{equation}
T_{ij}=\delta_{ij}-\widehat{n}_{i}\widehat{n}_{j},\label{eq: Tij}\end{equation}

and the longitudinal projector defined by

\begin{equation}
L_{ij}=\widehat{n}_{i}\widehat{n}_{j}\label{eq: Lij}\end{equation}

have been used. In this way the geodesic deviation equation (\ref{eq: deviazione geodetiche})
can be rewritten like

\begin{equation}
\frac{d^{2}}{dt^{2}}x_{i}=\frac{1}{16}\partial_{t}^{2}h_{g}T_{ij}x_{j}-\frac{1}{16}m^{2}h_{g}L_{ij}x_{j}.\label{eq: TL}\end{equation}

Thus it appears clear what was claimed in previous section: the effect
of the mass present in the GW generates a longitudinal force proportional
to $m^{2}$ which is in addition to the transverse one. But if $v(\omega)\rightarrow1$
in eq. (\ref{eq: velocita' di gruppo 2}) we get $m\rightarrow0$,
and the longitudinal force vanishes. Thus it is clear that the longitudinal
mode arises from the fact that the GW does no propagate at the speed
of light.

\section{Detectability of the polarization and angular dependence}

Now, let us analize the dectability of the polarization (\ref{eq: hg})
computing the pattern function of a detector to this massive component.
One has to recall that it is possible to associate to a detector a
\textit{detector tensor} that, for an interferometer with arms along
the $\hat{u}$ e $\hat{v}$ directions with respect the propagating
gravitational wave (see figure 1), is defined by \cite{key-2,key-13,key-14}\begin{equation}
D^{ij}\equiv\frac{1}{2}(\hat{v}^{i}\hat{v}^{j}-\hat{u}^{i}\hat{u}^{j}).\label{eq: definizione D}\end{equation}

\begin{figure}
\includegraphics{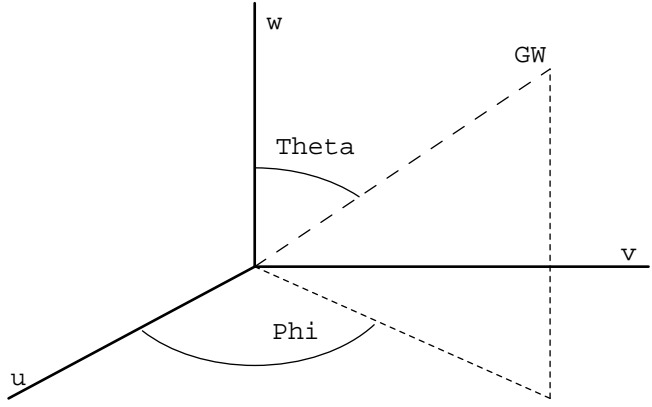}

\caption{a GW propagating from an arbitrary direction}

\end{figure}
 If the detector is an interferometer \cite{key-1,key-2,key-3,key-4,key-5,key-6,key-7,key-8,key-21},
the signal induced by a gravitational wave of a generic polarization,
here labelled with $s(t),$ is the phase shift, which is proportional
to \cite{key-2,key-21}

\begin{equation}
s(t)\sim D^{ij}\widetilde{R}_{i0j0}\label{eq: legame onda-output}\end{equation}

and, using equations (\ref{eq: eqs}), one gets

\begin{equation}
s(t)\sim-\sin^{2}\theta\cos2\phi.\label{eq: legame onda-output 2}\end{equation}

The angular dependence (\ref{eq: legame onda-output 2}), which is
shown in figure 2, %
\begin{figure}
\includegraphics{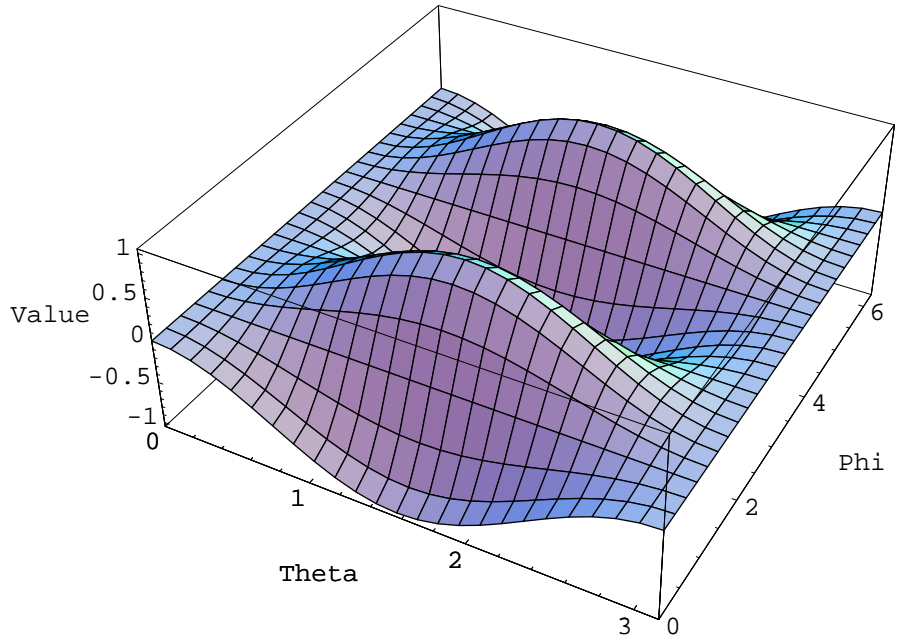}

\caption{the angular dependence (\ref{eq: legame onda-output 2}) }

\end{figure}
is different from the two well known standard ones arising from general
relativity which are, respectively $(1+\cos^{2}\theta)\cos2\phi$
for the $+$ polarization and $-\cos\theta\sin2\vartheta$ for the
$\times$ polarization. Thus, in principle, the angular dependence
(\ref{eq: legame onda-output 2}) could be used to discriminate among
the bimetric theory and general relativity, if present or future detectors
will achieve a high sensitivity.

\section{Conclusions}

After a brief review of the work of de Paula, Miranda and Marinho
on massive gravitational waves arising from a bimetric theory of gravity,
in this paper it has been shown that the presence of the mass generates
a longitudinal component in a particular polarization of the wave.
The effect of this polarization on test masses has been performed
using the geodesic deviation. A the end of this paper the detectability
of this particular polarization has also been discussed, showing that
its angular dependence could, in principle, discriminate such polarization
with respect the two ones of general relativity if present or future
detectors will achieve a high sensitivity.

As a final remark, it seems from the analysis in the last section
of this paper, that the investigation of massive component of gravitational
waves could constitute a further tool to discriminate among several
relativistic theories of gravity on the ground \cite{key-12,key-15,key-19,key-20}.

\section*{Acknowledgements }

The EGO consortium has to be thanked for the use of computing facilities

\end{document}